\documentclass[10pt,twocolumn]{article}
\usepackage{amsmath}
\usepackage{url}
\usepackage{graphicx}
\usepackage{color}
\usepackage{booktabs}
\usepackage[normalem]{ulem}
\usepackage{wasysym}
\usepackage{microtype}
\usepackage{wasysym}
\usepackage{xspace}
\usepackage{graphicx}
\usepackage{authblk}
\usepackage{enumitem}
\usepackage{verbatim}
\usepackage{tabularx}
\usepackage{amsmath}

\usepackage{epsfig,endnotes}

\usepackage{times}

\usepackage[small,compact]{titlesec}
\newcommand{\captionfonts}{\footnotesize}
\makeatletter
\g@addto@macro\normalsize{%
  \setlength\abovedisplayskip{3pt}
  \setlength\belowdisplayskip{3pt}
  \setlength\abovedisplayshortskip{3pt}
  \setlength\belowdisplayshortskip{3pt}
}
\makeatother

\makeatletter  
\long\def\@makecaption#1#2{%
  \vskip\abovecaptionskip
  \sbox\@tempboxa{{\captionfonts #1: #2}}%
  \ifdim \wd\@tempboxa >\hsize
    {\captionfonts #1: #2\par}
  \else
    \hbox to\hsize{\hfil\box\@tempboxa\hfil}%
  \fi
  \vskip\belowcaptionskip}
\makeatother   

\def\compactify{\itemsep=2pt \topsep=2pt \partopsep=1pt \parsep=1pt \leftmargin=1.6em}
\let\latexusecounter=\usecounter

\newcommand{\mxnet}{MXNet\xspace}
\newcommand{\phub}{PHub\xspace}

\newcommand{\code}[1]{\texttt{\small{#1}}}
\newcommand{\phubbox}{PBox\xspace}
\newcommand{\psliteib}{\mxnet IB\xspace}
\newcommand{\pshard}{PShard\xspace}

\newcommand{\mysection}{\S}

\begin{document}
	
\date{}
	
\title{Parameter Hub: a Rack-Scale Parameter Server for Distributed Deep Neural Network Training}

\author{}
\author{Liang Luo$^*$,  Jacob Nelson$^\dagger$, Luis Ceze$^*$, Amar Phanishayee$^\dagger$, Arvind Krishnamurthy$^*$ \\
$^*$University of Washington, $^\dagger$Microsoft Research}

\maketitle

	
\begin{abstract}

Distributed deep neural network (DDNN) training constitutes an increasingly important workload that frequently runs in the cloud. Larger DNN models and faster compute engines are shifting DDNN training bottlenecks from computation to communication.
This paper characterizes DDNN training to precisely pinpoint these bottlenecks. We found that timely training requires high performance parameter servers (PSs) with optimized network stacks and gradient processing pipelines, as well as server and network hardware with balanced computation and communication resources. We therefore propose \phub, a high performance multi-tenant, rack-scale PS design. \phub co-designs the PS software and hardware to accelerate rack-level and hierarchical cross-rack parameter exchange, with an API compatible with many DDNN training frameworks. \phub provides a performance improvement of up to 2.7x compared to state-of-the-art distributed training techniques for cloud-based ImageNet workloads, with 25\% better throughput per dollar.


\end{abstract}

\section{Introduction}
\label{sec:introduction}
To date, most work in the systems and architecture community has focused on improving the efficiency of evaluating trained models. However, arriving at a trained model frequently requires experimentation, and thus multiple training runs, each of which may take days. Accelerating the training process lets neural network developers iterate faster and design better models.

As DNNs get computationally more expensive to train, timely training requires exploiting parallelism with a distributed system, especially in the cloud~\cite{CloudTPU88:online,MachineL65:online,ApacheMX19:online}. The most common way of exploiting parallelism is called ``data'' parallelism, which consists of a computational-heavy forward and backward phase and a communication-heavy parameter exchange phase.

In this paper, we begin by performing a detailed bottleneck analysis of DDNN training and observe that the emergence of speedier accelerators shifts \textit{the performance bottleneck of distributed DNN training from computation to communication}, because of the following factors. 

\begin{figure}[t!]
	\includegraphics[width=\linewidth,trim=3 13 3 3,clip]{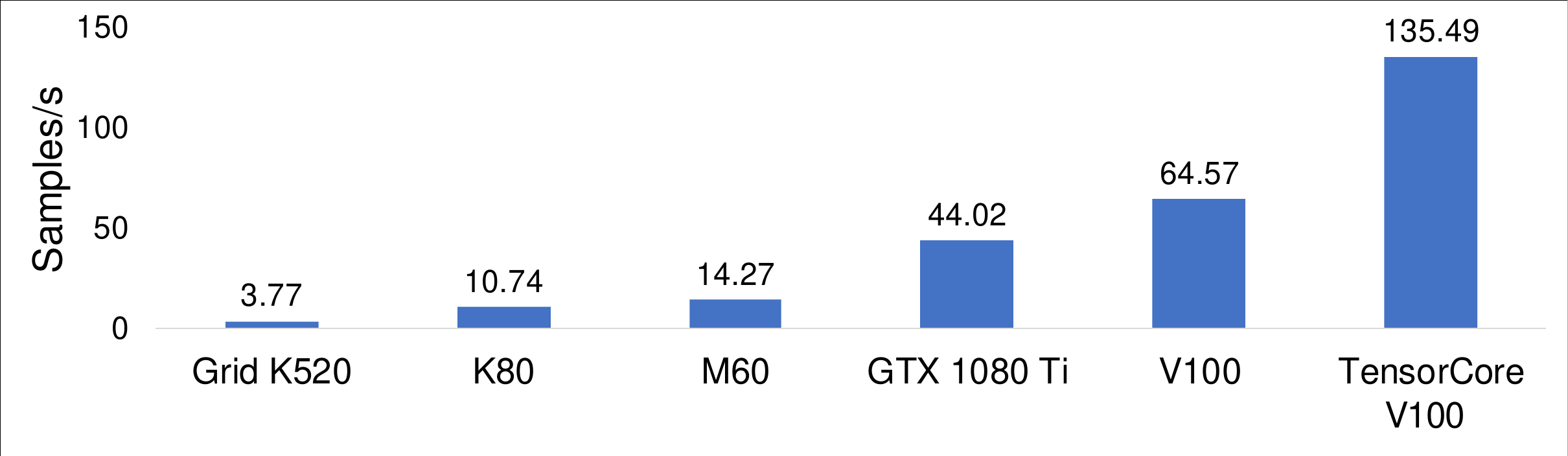}
	\caption{Single-GPU training throughput for ResNet 269 measured with \mxnet on EC2 g2, p2, g3 and p3 instances, and a local GTX 1080 Ti machine, while maximizing GPU memory utilization. Per chip GPU throughput on ResNet 269 in the cloud has increased 35x since 2012.}
	\label{fig:GPUPower}
\end{figure}

First, the throughput of GPUs on a recent DNN, ResNet, has increased by 35x since 2012 on modern cloud-based GPUs (Figure~\ref{fig:GPUPower}), effectively demanding a similar increase in network bandwidth, causing bandwidth deficiency.
Upgrading datacenter networks is expensive: compute instance network bandwidth on major cloud providers such as EC2 has improved little across generational upgrades~\cite{ec2BW}, so care must be taken when configuring racks for DDNN training for optimal cost-efficiency.

Second, existing parameter exchange mechanisms such as parameter servers (PS) do not scale up the total throughput on a standard cloud network stack~(Table~\ref{table:frameworkPerf}) due to unoptimized software and hardware configuration, and lack of awareness of the underlying physical network topology. 

\begin{figure}[t!]
	\includegraphics[width=\linewidth,trim=4 13 3 4,clip]{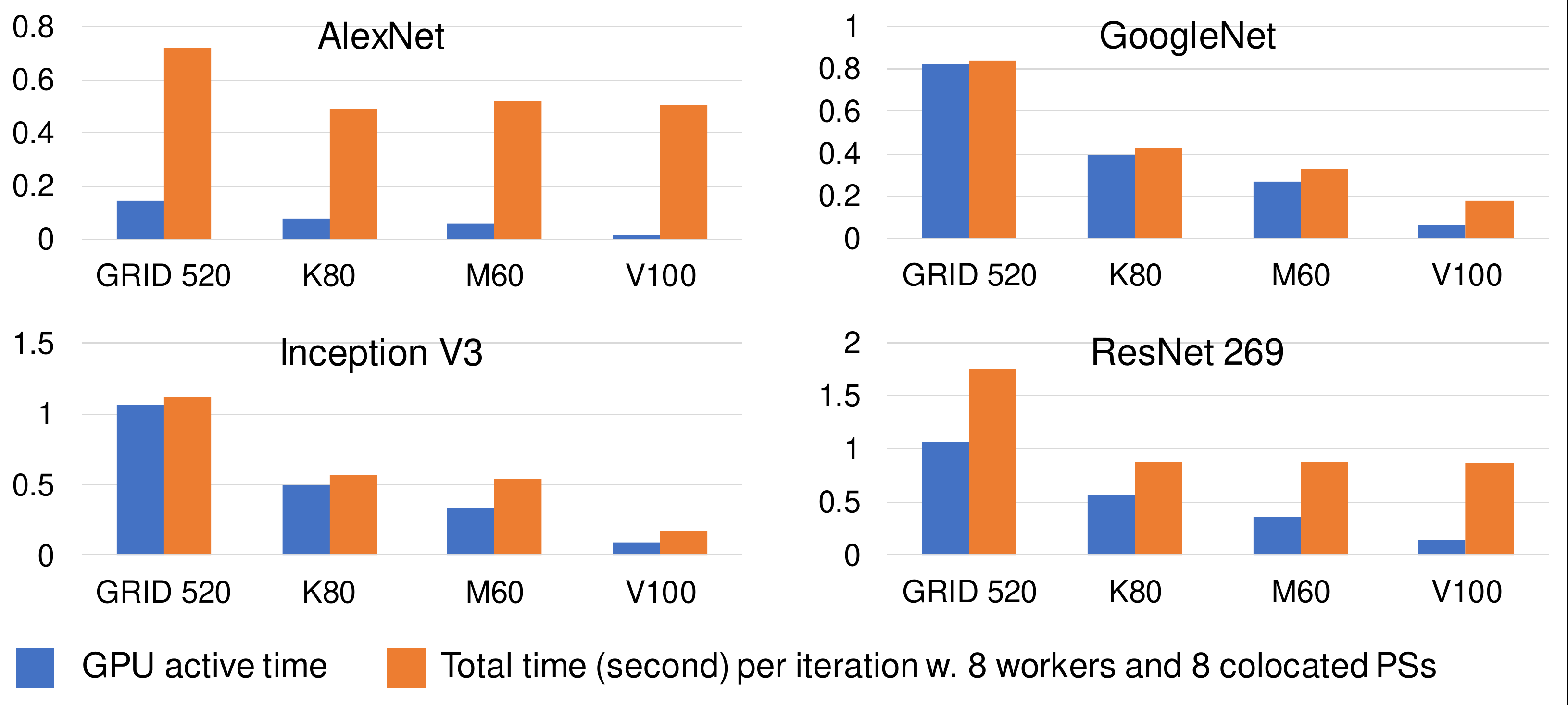}
	\caption{The overhead of distributed training gets larger as GPUs get faster. The framework can no longer hide communication latency, and faster GPUs no longer improve training throughput. With today's fast GPUs, distributed cloud DNN training time is chiefly spent waiting for parameter exchanges.}
	\label{fig:IncreasingCommToCompRatio}
\end{figure}

The compound effect of these factors dramatically increases communication overhead during distributed DNN training. To illustrate this problem, Figure \ref{fig:IncreasingCommToCompRatio} shows a modest-scale DNN training with 8 machines on EC2 with 10 Gbps links\footnote{
Batch size per GPU (4, 16, 32, 32, saturating GRID 520) for each network is kept the same across all GPUs for easier comparison.}: modern DNN training frameworks can no longer hide the latency of communication due to faster computation. Spending most of the DDNN training time on model updates limits the benefit of faster GPUs.

Scaling cloud-based DDNN training throughput demands both a fast and a cost-effective solution. Our bottleneck findings show that such a solution requires a more optimized software stack, a specialized hardware design, and a more effective cluster configuration.

We therefore propose \phub, a high performance, multi-tenant, rack-scale PS design for cloud-based DDNN training. By co-designing the PS software with the hardware and the datacenter cluster rack configuration, \phub achieves up to 2.7x faster training throughput, with 25\% better throughput per dollar.
Our contributions include:
\begin{enumerate}[noitemsep,topsep=0pt,parsep=0pt,partopsep=0pt]
	\item A detailed bottleneck analysis of current state-of-the-art cloud-based DDNN training~(\mysection\ref{sec:background}).
	\item Design and implementation of the \phub PS software, supporting many DNN training frameworks~(\mysection\ref{sec:design}).
	\item A \textit{balanced} central PS hardware architecture, \phubbox(\mysection\ref{sec:phub}), to leverage \phub for rack-level and hierarchical cross-rack gradient reduction.
	\item A comprehensive evaluation of \phub in terms of performance, scalability, and deployment cost~(\mysection\ref{sec:evaluation}).
\end{enumerate}

\section{Bottlenecks in Cloud-Based Training}
\label{sec:background}

\begin{figure}
	\includegraphics[width=\linewidth,trim=4 4 4 4,clip]{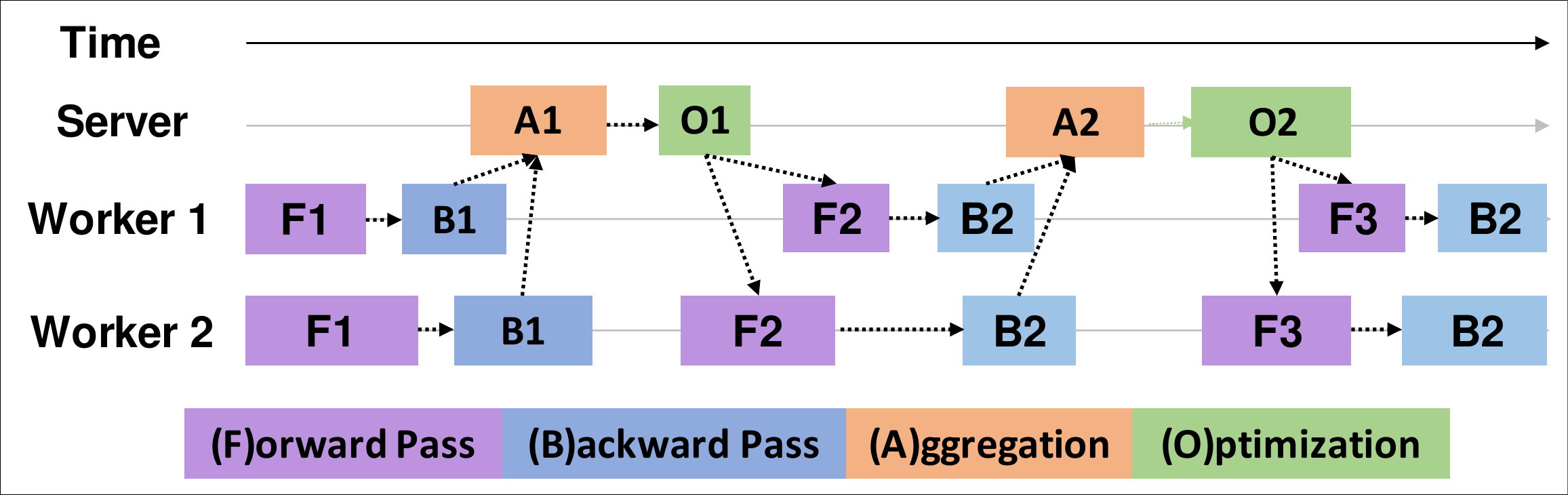}
	\caption{A few iterations in distributed training.}
	\label{fig:ddnn}
\end{figure}




Modern neural networks can have hundreds of \textit{layers} making up multi-megabyte-size \textit{models}.
The training process has three phases. In the \textit{forward pass}, a prediction is generated for an input. In the \textit{backward pass}, the prediction is compared with a label to calculate prediction error; then, through \textit{backpropagation}~\cite{backprop}, the gradient for each parameter 
is calculated with respect to this error. The model is then \textit{updated} using these gradients, often using a variant of the gradient descent optimization algorithm. Computation is often done on GPUs or other accelerators suited to regular data-parallel operations, processing tens to hundreds of samples at once (\emph{minibatching}).

The distributed training process (Figure~\ref
{fig:ddnn}) 
is different in a few ways.
First, a mean gradient is calculated across all minibatches in all the GPUs in each machine. Then, the mean of the gradients from each machine is calculated. Finally, the model is updated based on that mean, new parameters are broadcast to each machine and GPU, and the next batch is trained. This paper focuses on optimizing calculation for both the mean gradient across machines and subsequent model updates (or \textit{parameter exchange}).

In a typical DDNN training setup, machines can take the role of a worker and/or a parameter server (PS). PSs are specialized key-values stores that collect the gradients and update the model~\cite{ps0,ps1,ps2,ps3}. In this paper, we use ``key'' to refer to a layer, and ``value'' to refer to the set of parameters for that layer. 


The process described here is \emph{synchronous training}, where all machines and GPUs execute a new minibatch simultaneously and update the model based on the gradients in the current iteration. It is also possible to train asynchronously~\cite{tensorflow,revisitSGD,GeePS,recht2011hogwild,projectAdam,googleDNN} or with relaxed consistency~\cite{DBLP:journals/corr/DaiKWHGX14,SSP,BSP,184014,Wei:2015:MCC:2806777.2806778}, sacrificing reproducibility for potential throughput increase. 
We focus on synchronous training due to its simplicity and commonality in industry, but our techniques can also benefit asynchronous training.

\begin{figure}[tb!]
	\includegraphics[width=\linewidth,trim=1 5 1 0,clip]{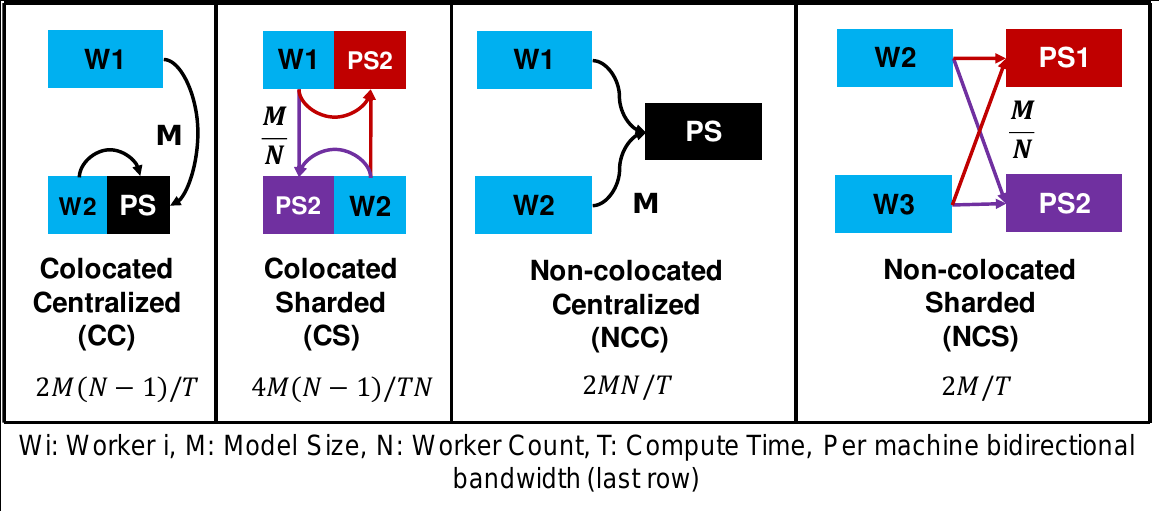}
	\caption{PS configurations in a DDNN training system, and minimum network bandwidth to fully hide communication overhead.}
	\label{fig:pssetups}
\end{figure}

\subsection{Common PS Configurations}
PS configurations primarily differ along two axes: colocated (C) versus non-colocated (NC), and centralized (C) versus sharded (S). 
A PS setup is colocated if a worker and a server process share the same physical machine. A PS setup is centralized if a single PS process handles all keys; and a sharded setup load-balances keys across multiple PS processes.
During synchronization, each worker sends and receives model updates from each PS process. Figure \ref{fig:pssetups} illustrates the four combinations of choices from these two axes: Colocated Centralized (CC), Colocated Sharded (CS), Non-colocated Centralized (NC) and Non-colocated Sharded (NCS).

In general, sharded PSs scale better at higher hardware costs. Colocated PSs reduce total data movement on the network by $\frac{1}{N}$ with $N$ workers participating: the update for the partition of the model assigned to a colocated PS need not go through the network. While many frameworks default to CS configurations~\cite{MXNetont0:online, Distribu25:online}, in a colocated setup the PS process interferes with the training process, because both are contending for network and processing resources. Specifically, compared to NC PSs, \textit{each network interface must process roughly 2x the network traffic, because both the colocated worker and PS processes must send and receive model updates from remote hosts}, creating a major bottleneck in network-bound DDNN training. 

\subsection{The \mxnet Framework}

\mxnet\cite{chen2015mxnet} is a widely used, state-of-the-art DDNN training framework that supports many new optimizations in the literature~\cite{poseidon,chen2016training,GeePS}. It is widely used on AWS~\cite{ApacheMX19:online}, and natively supports distributed training: its PS implementation relies on TCP, built on top of the ZMQ~\cite{zmq} distributed messaging library.

All modern DNN training frameworks can fully utilize GPU resources by taking advantage of primitives, such as CuDNN. These frameworks offer comparable throughput when training DNNs. For distributed training, many frameworks such as \mxnet provide eager scheduling of parameter exchanges, overlapping backward computation with parameter synchronization, hiding communication latency. We measured distributed training performance and scalability for Caffe2, TensorFlow, and \mxnet{}\footnote{Caffe2: Halving and doubling. TensorFlow and \mxnet: CS PSs with a 1:1 worker-to-PS ratio. Network: 56 Gbps IPoIB. GPU: GTX 1080 Ti. Neural Network: ResNet 50 with batch size of 32. Poseidon hangs when more than 5 workers training this network in our cluster. 8-worker throughput is overestimated as per worker throughput (at 5 workers) * 8.} 
with up to 8 worker nodes. We found comparable throughput when training ResNet 50 on a 56 Gbps network using SGD, with \mxnet slightly leading the pack (Table \ref{table:frameworkPerf}). These results align well with other observations \cite{Chainer,1608.07249,zhudnn}. Therefore, we use \mxnet as the basis for our implementations and optimizations, but the techniques are generalizable.

\begin{table}[tb!]
        \centering
        \footnotesize
	\begin{tabular}{|c|c|c|c|c|}
		\hline
		   & Local & 2 nodes & 4 nodes & 8 nodes\\
		\hline 
		TensorFlow   & 152  &  213  & 410    &  634 \\
		\hline
		Caffe2 & 195   &  266  &  343   &   513  \\
		\hline
		TF+Poseidon\cite{poseidon} & 209 & 229  & 364 & $<$648 \\
		\hline
		\mxnet & 190  &  187  &  375   &  \textbf{688}  \\
		\hline
	\end{tabular}
	\caption{Throughput (samples/s) of major DNN training frameworks with a 56 Gbps network.}
	\label{table:frameworkPerf}
\end{table}


\subsection{Bottleneck Findings}
\label{sec:cloudTraining}
Ideally, communication latency is fully hidden by computation, i.e., compute engines never wait for data. In reality, since computation speed exceeds communication speed in cloud-based DDNN training, time is wasted waiting for model updates (Figure \ref{fig:IncreasingCommToCompRatio}). Workers run much faster locally~(Table \ref{table:frameworkPerf}), so the bottlenecks must lie in the PS, the network stack, and/or the physical network itself. Our study finds three major bottlenecks in cloud-based DDNN training: insufficient network bandwidth, framework inefficiencies, and suboptimal deployment in the cluster. We elaborate on each below. 

\subsubsection{Insufficient Bandwidth}
We profiled the training of multiple DNNs of different model sizes and computation-to-communication ratios. 
Our setup used 8 workers and 8 CS PSs.
We observed \textit{it was nearly impossible to eliminate communication latency in cloud-based training due to limited network bandwidth.}
We estimated the minimum bandwidth requirement to fully hide communication latency in the network as follows: given a model size of $M$, and $T$ time for each iteration, with $N$ workers participating, the network should at least be able to send and receive model updates within the computation time (assuming infinitely fast PSs and that sending/receiving could fully overlap). Figure \ref{fig:pssetups} gives an analytical lower bound of \textit{per host bandwidth}, and Table \ref{table:bwReqDC} shows the required bandwidth for various DNNs: DNNs demand more bandwidth than mainstream cloud providers offer (typically 10-25 Gbps) in the VMs. 
A similar observation was made in prior work~\cite{alannetwork, zhudnn}. Furthermore, bandwidth requirement increases with worker count.

\subsubsection{Framework Bottlenecks}
\begin{figure}[t!]
	\includegraphics[width=\linewidth,trim=3 2 2 4,clip]{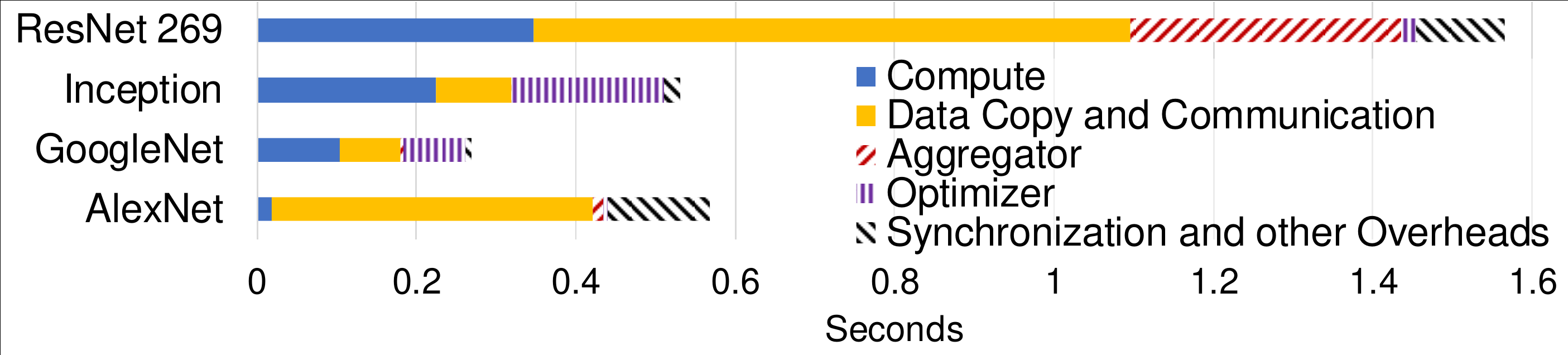}
	\caption{Progressive overhead breakdown of different stages during the distributed training pipeline for \mxnet DDNN training on a 56Gbps network. Link capacity accounts for a small fraction of the copy and communication overhead in this setting.}
	\label{fig:overheadBreakdown}
\end{figure}

However, even with ample communication resources, existing PSs failed to hide communication latency and struggled to scale. Table \ref{table:frameworkPerf} shows that all major DNN training frameworks 
do not scale well with a 56 Gbps IPoIB network.
We investigated the cause for \mxnet by breaking down the overhead for each major component of a training iteration (legends of Figure \ref{fig:overheadBreakdown}). Since all stages overlap one another, and since ideally we would like early stages to fully hide the latency of later stages, we show \textit{progressive overhead} in Figure \ref{fig:overheadBreakdown}: we gradually turned on different components in the \mxnet DDNN training pipeline, and each segment shows the \textit{additional overhead that previous stages could not hide}. Specifically, the compute segment shows how long the GPU is active; the data copy segment shows the additional overhead of turning on distributed training without aggregation and optimization; the aggregation and optimization segments show additional overheads of enabling them in that order; and the ``other'' overheads segment includes synchronization and overheads that are not specific to a single component. We explain the high overhead for some components:

\begin{table}
        \centering
        \footnotesize
	\begin{tabular}{|c|c|c|c|c|}
		\hline
		Network   & CC & CS & NCC & NCS\\
		\hline
		ResNet 269    & 122   & 31   & 140    &  17   \\
		\hline
		Inception & 44   &  11  &  50   &  6  \\
		\hline
		GoogleNet & 40   &  10  &  46   &  6  \\
	
		\hline 
		AlexNet   & 1232  &  308  & 1408    &  176  \\
		\hline
	\end{tabular}
	\caption{Estimated bisection bandwidth (Gbps) lower bound on the PS side for hiding communication latency. Same setup as Table \ref{table:frameworkPerf}.}
	\label{table:bwReqDC}
\end{table}

	\noindent\textbf{Data copy:} each layer's parameters were copied to and from OS buffers 4 times during parameter exchange.
\vspace{-0.1ex}	

	\noindent \textbf{Aggregation and optimization:} \mxnet's ``wide'' approach to achieving parallelism did not achieve high throughput in our measurements (see \mysection\ref{sec:tallvswide}). 
\vspace{-0.1ex}	
	
	\noindent \textbf{Synchronization:} \mxnet's dispatcher thread needs to synchronize access with ZMQ threads, aggregation threads and a optimization thread via shared queues, leading to bad locality and increased synchronization overhead. 



\begin{figure}[t!]
	\centering
	\includegraphics[width=.7\linewidth,trim=2 1 1 2,clip]{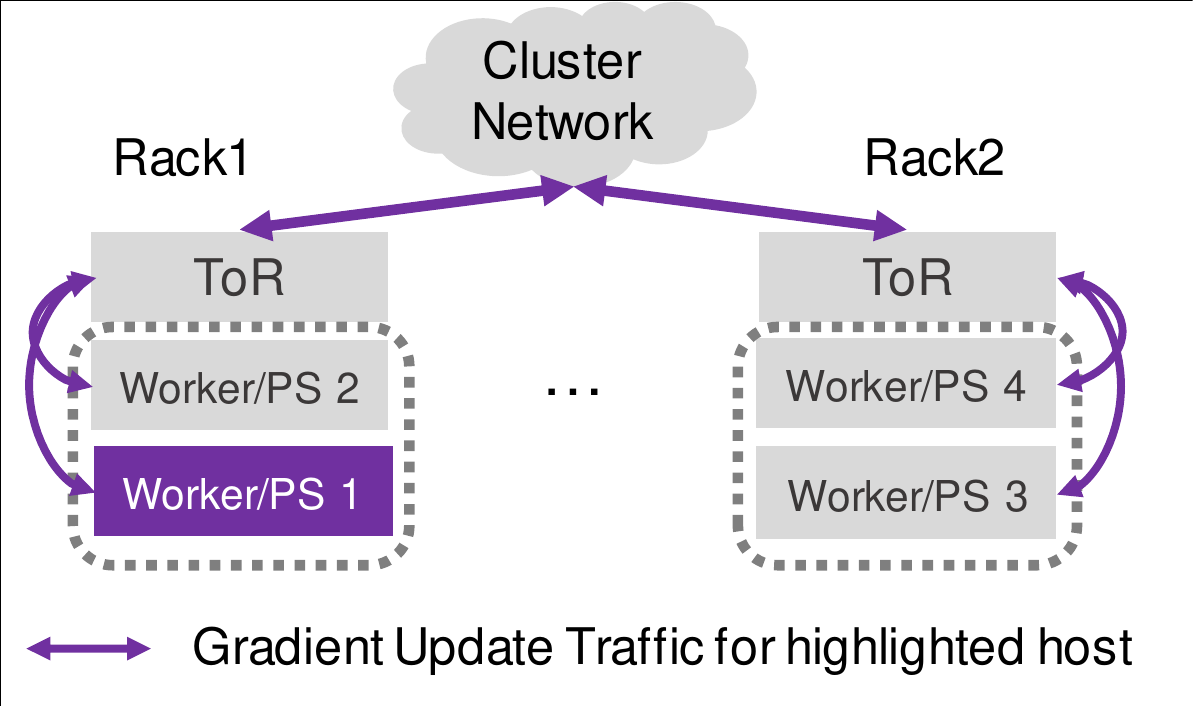}
	\caption{When workers/servers span multiple racks in cloud-based training, large delay occurs due to gradient update traffic going through an oversubscribed network core.}
	\label{fig:deploymentOverhead}
\end{figure}


\subsubsection{Deployment-related Overhead}
VMs associated with a training job can be far away from each other when launched in the cloud. Existing frameworks assume homogeneous inter-VM bandwidths, which results in excessive communications between distant VMs, in turn leading to bottlenecks.
We conducted an experiment to probe pair-wise bandwidth of 8 P2.8xLarge 10 Gbps instances on EC2 and found that bandwidths can vary by a factor of 2---even between send and receive streams of the same instance pair.
Some cloud environments support co-scheduling constraints (e.g., EC2 placement groups), but for large jobs on busy clusters it may take a long time to satisfy these constraints.


One possible reason is the oversubscribed network topology in the data center~\cite{Mysore2009PortLandAS,Roy2015InsideTS, incbricks}, providing full bisection bandwidth within each rack but not across racks when the core is busy. Thus, gradient update streams that go through a potentially oversubscribed network core~\cite{vl2-a-scalable-and-flexible-data-center-network,43837} can be delayed. Network topology awareness is crucial for DDNN workloads~\cite{DBLP:journals/corr/abs-1712-03890, wang2018optimal}.
In our work, we pursue a rack-scale PS that takes advantage of intra-rack full bisection bandwidth and minimizes inter-rack traffic via \textit{hierarchical reduction algorithms} (see Section \ref{sec:hierarchicalReduction}).

\section{\phub Design}
\label{sec:design}
Based on \mysection\ref{sec:cloudTraining} findings, we propose a rack-scale PS, \phub, that reduces framework overhead with software optimizations, mitigates bandwidth insufficiency with a re-architected, balanced server configuration, and lowers network environment-induced overhead with topology-aware reduction algorithms. With \phub, we aim to:


\begin{enumerate}[noitemsep,topsep=0pt,parsep=0pt,partopsep=0pt]
\item Minimize gradient/model communication overhead.
\item Enable efficient gradient processing and overlapping with communication.
\item Balance communication and computation capabilities, both within and PS and between workers and the PS.
\item Allow low interference of co-running jobs and minimized cross-rack traffic.
\end{enumerate}

%

\subsection{The \phub Service API and Interoperability with other Frameworks}
\phub{}'s API is designed for compatibility with multiple DNN training frameworks. Workers use \phub by first calling \code{PHub::CreateService} on the connection manager. This sets up access control and a namespace for the training job and returns a handle. The client side uses the handle to finish setup. \phub uses the namespace and an associated nonce for isolation and access control. 

Jobs call \code{PHub::ConnectService} to rendezvous servers and workers, exchanging addresses for communication. This call replaces \code{Van::Connect} in \mxnet, \code{Context::connectFullMesh} in Caffe2 and \code{GrpcServer::Init} in TensorFlow. \code{PHub::InitService} causes the current \phub instance to allocate and register receive and merge buffers. \phub also authenticates each worker's identity using the nonce. Authentication is a one-time overhead and once a connection is established, \phub assumes the remote identity associated with that address/port/queue number does not change during training.

\phub{}'s functional APIs include standard synchronous or asynchronous \code{PHub::Push/Pull} operations that are used in TensorFlow (\code{GraphMgr::SendInputs/RecvOutputs}) and \mxnet (\code{KVStoreDist::PushImpl/PullImpl}). \phub also includes a fused \code{PHub::PushPull} operation that perform a push, waits until all pushes are complete, and pulls the latest model. The fused operation often saves a network round-trip as push and pulls are frequently issued consecutively. This operator can serve as a drop-in replacement for Caffe2's \code{Algorithm::Run}.

\subsection{\phub Software Optimizations}
\label{sec:commonOptimizations}
This section describes software optimizations that benefit different stages in DDNN training across all common PS configurations. 

\subsubsection{Network Stack Optimizations}
\label{sec:IBOptimization}
We sought to mitigate data movement latency with zero-copy and kernel bypass. We chose InfiniBand (IB) since we were already familiar with the Verbs API, and it is available in major cloud providers~\cite{AzureWin5:online}. Note that similar results could be achieved over Ethernet using RoCE, DPDK or other frameworks. We followed the guidelines from~\cite{rdma}; we tried two and one-sided RDMA, and two-sided send/receive operations and found similar performance in our workload. We briefly highlight some implementation details:

\noindent \textbf{Minimal Copy:} Leveraging InfiniBand's zero-copy capability, the only required data copy is between the GPU and main memory. When one GPU is used, this can be eliminated with GPU-Direct RDMA on supported devices.

\noindent \textbf{NUMA-Aware, One-shot Memory Region Registration:} Since a worker can operate on only one model update at a time, it is sufficient to allocate one read buffer (for the current model) and one write buffer (for update reception) for the model. To minimize InfiniBand cache misses, \phub preallocates all buffers in the NUMA domain where the card resides as a contiguous block.

\noindent \textbf{Minimal Metadata:} To maximize bandwidth utilization and minimize parsing overhead, \phub encodes metadata (such as callback ID and message opcode) into InfiniBand's queue pair number and immediate field. This saves \phub an additional PCIe round trip (from IB send scatter/gather) to gather metadata when sending messages.


\subsubsection{Gradient Aggregation and Optimization}
\label{sec:tallvswide}
Gradient aggregation and optimization are element-wise operations. Aggregation sums gradients for the same key from all workers. Optimization updates the model using aggregated gradients with an algorithm such as SGD. Our design goal is to overlap aggregation and optimization of different keys with communication.

Gradient aggregation could occur in the CPU or GPU~\cite{GeePS}. Here, we posit that the CPU is sufficient for this job.
Aggregation is simply vector addition: we read two floats and write one back.
With our typical modern dual socket server, if we keep our processors' AVX ALUs fed, we can perform 470 single-precision giga-adds per second, requiring 5.6 TB/s of load/store bandwidth.
But the processors can sustain only 120 GB/s of DRAM bandwidth. 5.6 TB/s is impractical 
in DNN training workloads, making aggregation inherently memory bound. Thus, copying gradients to a GPU for aggregation is not helpful.

\begin{figure}[t!]
	\includegraphics[width=\linewidth,trim=5 5 15 5,clip]{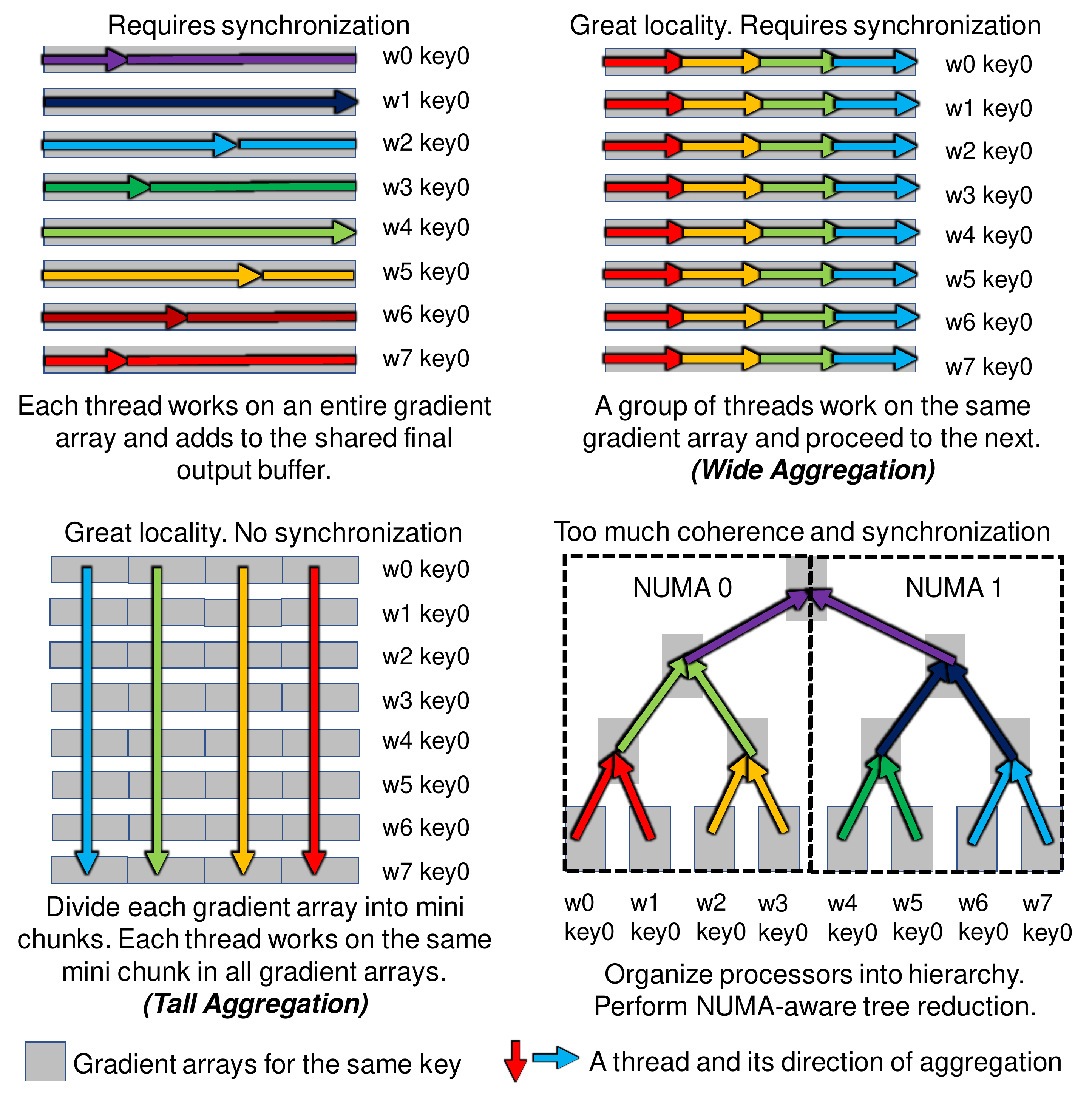}
	\caption{Ways of gradient aggregation. A thread (arrow) aggregates over the array (gray rectangle) of gradients from a worker. }
	\label{fig:gradientAggregation}
\end{figure}

There are many ways to organize threads to perform aggregation.
Figure \ref{fig:gradientAggregation} shows four options we prototyped， assuming gradient arrays are available at once. We found that the best performance was achieved using the two discussed below; other schemes suffered from too much synchronization overhead, poor locality and/or high latency.

\textit{Wide aggregation} is typical to systems like \mxnet that call BLAS routines for linear algebra. In these systems, a group of aggregation threads process one gradient array at a time; each thread works on a partition of that array. 

A variation of wide aggregation is \textit{tall aggregation}, which chunks a gradient array into mini-chunks of predefined sizes; each thread works independently to process the same chunk across all gradient arrays for a given key. This is the preferable way to organize threads for many reasons.
First, gradient arrays do not arrive instantly. For a large key (e.g., a fully connected layer), aggregation and optimization cannot start for wide aggregation until the key is fully received; for tall aggregation, the process can start as soon as the first chunk is received.
Second, in wide aggregation, it is challenging to balance the number of threads dedicated to aggregation and to optimization, let alone partitioning threads to work on different keys since they can arrive at the same time; 
thread assignment for tall aggregation is natural.
Third, wide aggregation induces queuing delays: it effectively processes one key at a time versus tall aggregation's many ``mini-queues.''
Fourth, wide aggregation puts many threads to work in lock-step on pieces of data, which incurs non-trivial synchronization overhead; tall aggregation requires no coordination of threads as aggregation is an element-wise operation.



\phub tracks the number of currently aggregated mini-chunks for a given key. When a chunk is received from all workers, it can be optimized. This step is natural in \phub: the thread that aggregates a particular chunk also optimizes that chunk. As a result, \phub{}'s aggregation and optimization scheme effectively maps a particular chunk to a single core (since \phub pins threads to cores). On the other hand, \mxnet uses wide optimization: when a key is fully aggregated, another set of threads is launched to perform aggregation. No overlap occurs between key aggregation and optimization. 

We explored the benefits of caching by implementing two variants of each aggregator and optimizer: one using normal cached loads and stores, and one with non-temporal prefetches and stores. We found it beneficial to cache both the model and gradients. \phub{}'s aggregators and optimizers are fully extensible: implementations that comply with \phub{}'s API can be used during runtime.

\subsubsection{Fine-grained Key Chunking}


Chunking in \phub differs from chunking in other systems in key ways. Initially, our goal is to balance load at a fine-grained level across cores and interfaces rather than across server shards: chunking is turned on even when a centralized PS is used. Next, we would expect our optimal chunk size to be the nearest, smallest message size that can saturate network bandwidth, whereas systems like \mxnet prefer larger key chunk sizes to avoid excessive thread synchronization overhead.
In fact, \phub's default is 32KB, while \mxnet's is 4MB.
Finally, key chunking enables another important optimization: the overlapping of gradient transmission with aggregation and optimization. Aggregation starts only after a key's entire gradient array is received; and for large layers, this adds significant delay. With small key chunks, \phub enables ``streaming'' aggregation and optimization.

\begin{figure}[t!]
	\centering
	\includegraphics[width=.8\linewidth,trim=1 0 1 1,clip]{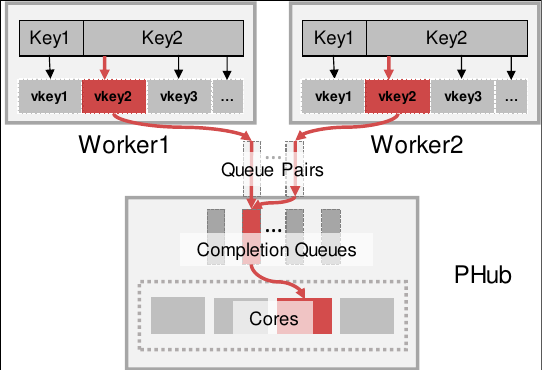}
	\caption{The process of mapping a chunk to a core in \phub using fine grained key chunking. Keys are chunked into virtual keys. The highlighted key is delivered to a highlighted (fixed) core through a highlighted (fixed) queue pair and completion queue. }
	\label{fig:mappingChunkToCore}
\end{figure}

\subsubsection{Mapping a Chunk to a Core}
\label{sec:chunk2core}
\phub's assignment of chunks to cores is computed during initialization. At that time, the set of all keys is sharded across the cores and interfaces available on PS nodes.
A specific chunk is always directed to a particular queue pair,
which is associated with a shared completion queue on the chunk's core.
All message transmission, reception, and processing for that chunk is done on that core. Cores do not synchronize with each other. Once processed, a chunk is transmitted back to the workers on its originating path. The worker side of \phub assembles and disassembles a key, a process that is transparent to the framework.

\phub{}'s chunk assignment scheme provides significant locality benefits. The same key likely arrives around the same time from multiple workers; the associated aggregation buffer is reused during this period. The scheme also encourages locality in the InfiniBand interface in the queue pair and memory registration caches, which can further benefit performance.

This scheme imposes challenges in balancing load across cores, queue pairs and completion queues. \phub uses a 4/3 approximation set partition algorithm to balance each component's workload at each level, which produces practically balanced assignments in our experiments.

\subsection{A Balanced Hardware Design for Rack-Scale PSs}
\label{sec:phub}
Centralized PSs have lower cost than NCS PSs, and half of the bandwidth stress compared to CS PSs on each interface card. Thus it is desirable to have a centralized reduction entity at rack level. However, scaling a centralized PS to rack scale is challenging~\cite{firecaffe}, despite the optimizations in \mysection\ref{sec:commonOptimizations}.
The root cause is hardware imbalance in allocation of computation and communication resources in the host machine: centralized PSs usually run on the same hardware configuration as a worker, which have only one or two network interfaces. This implies incast congestion from their high bandwidth usage (Table \ref{table:bwReqDC}) when serving multiple workers, starving the compute units. 



One trivial solution would be to simply use interfaces with higher bandwidth. However, even in the best case, a single network interface is not capable of saturating memory or PCIe bandwidth. A single network interface also causes serialization delay and further imbalance across NUMA domains in a typical server. 



\begin{figure}[t!]
	\centering
	\includegraphics[width=.8\linewidth,trim=1 0 1 1,clip]{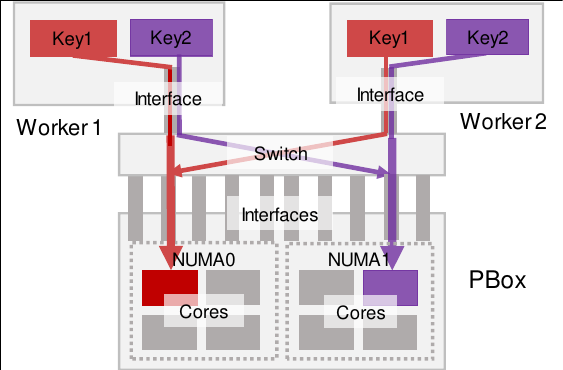}
	\caption{The \phubbox architecture}
	\label{fig:phub}
\end{figure}

This section describes \phubbox, our \textit{balanced parameter exchange system}. We maintain that a centralized system, when architected properly, can provide high throughput, low latency, sufficient scalability for a rack, and low cost. We prototyped \phubbox using an off-the-shelf server platform that was configured to our requirements. Our goal was to balance IO and memory bandwidth; our prototype system had memory bandwidth of 120 GB/s and theoretical overall bidirectional IO bandwidth of 140 GB/s. To fully utilize resources, \phubbox needed a matching network capability, which we provided by using multiple network interfaces. Figure \ref{fig:phub} shows the resulting \phubbox design. The system includes 10 network interfaces, each of 56 Gbps link speed, connected to a switch. This uses all PCIe bandwidth on our dual socket prototype and provides roughly 136 GB/s bandwidth once IB and PCIe framing overheads are taken into account, balancing IO and memory bandwidth.

Hardware alone solves only part of the problem. Existing frameworks cannot efficiently use the full hardware capability even if multiple interfaces are present (for example, TensorFlow and \mxnet support multiple interfaces only by spawning multiple PS processes). Thus, software that understands both the hardware topology \textit{and} balance is required to complete the solution. \phub takes full advantage of \phubbox by extending the chunk-to-core mapping scheme (\mysection\ref{sec:chunk2core})
, ensuring balance across interfaces and NUMA domains. \phub further guarantees no inter-processor traffic on \phubbox, and completion queues and queue pairs in an interface card are used by only one core in the same NUMA domain to promote locality and avoid coherence traffic. In essence, \phubbox forms micro-shards inside a box.


\subsection{Rack Deployment and Topology-Aware Reduction}
\label{sec:hierarchicalReduction}

\begin{figure}[t!]
	\centering
	\includegraphics[width=.7\linewidth,trim=2 1 1 2,clip]{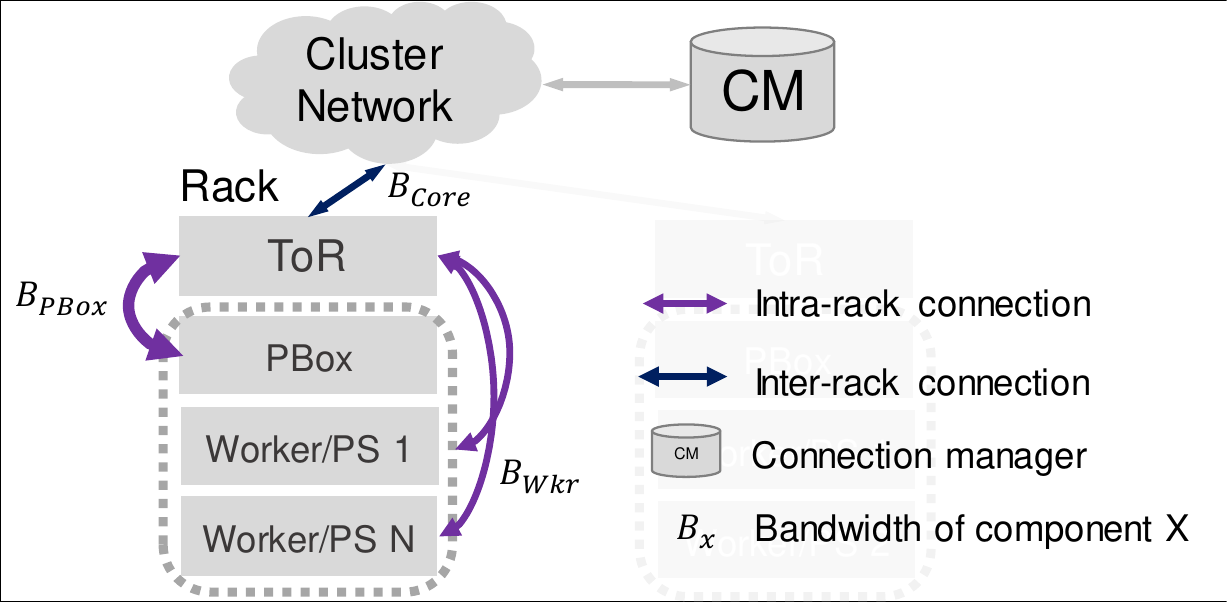}
	\caption{\phubbox deployment scheme}
	\label{fig:pBoxDeployment}
\end{figure}

We associate a \phubbox with a ToR during deployment for two reasons. First, full bisection bandwidth is achievable for machines in the same rack, making it ideal for a central reduction entity as \phubbox, while oversubscription occurs between the ToR and the cluster network. Second, as we show in \mysection\ref{sec:scability}, a single \phubbox has enough scalability for a typical rack of worker machines.

When provisioned in each rack (Figure \ref{fig:pBoxDeployment}), \phubbox{}es can form an array of sharded PSs, or run a \textit{hierarchical reduction} algorithm for a training task that spans multiple racks through the coordination of a connection manager. Hierarchical reduction works in three steps: first, each \phubbox centrally aggregates gradient updates from workers in the same rack; then, the \phubbox{} nodes start cross-rack aggregation and compute globally aggregated gradients; finally, each per-rack \phubbox runs an optimizer on this gradient and broadcasts the new weights back to local workers.

Hierarchical reduction trades off more rounds of communication for lower cross-rack traffic ($1/N$ with N-worker racks). \phub determines when hierarchical reduction is potentially beneficial with the simple model below:
\[	max(\frac{N-1}{B_{bn}}, \frac{1}{NB_{Wkr}}) > max(\frac{1}{B_{PBox}}, \frac{N}{B_{Wkr}}) + C\]

where $B_{PBox}, B_{Core}$ and $B_{Wkr}$ are the bandwidths of a \phubbox, the network core, and a worker,  $B_{bn} = min((r-1)B_{PBox}, B_{Core})$, and $r$ is the number of racks. When the condition is true, this means the time to perform cross-rack transfer is larger than the added latency of a two-level reduction, which consists of a per-rack local aggregation that happens in parallel and an inter-rack communication (with cost $C$) done with either sharded PSs ($C=\frac{N-1}{NB_{bn}}$) or a collectives operation (e.g., $C\approx\frac{r-1}{rB_{bn}}$ with racks forming a ring). \mysection\ref{sec:hierarchicalEval} estimates the overhead of $C$, and $B_{Core}$ can be effectively probed by using~\cite{hu2002estimating,hu2003evaluation}.


\section{Evaluation}
\label{sec:evaluation}
We added support for \phub{}'s API to \mxnet, replacing its PS. We evaluated \phub by comparing it to \mxnet{}'s unmodified PS. We had four goals in our evaluation: (1) to assess the impact of \phub software and the \phubbox hardware on training throughput, (2) to identify the importance of each optimization, (3) to determine the limits of \phubbox, (4) to evaluate effectiveness of \phubbox as a rack-scale service. and (5) to demonstrate the cost-effectiveness of the \phub.
\subsection{Experimental Setup}
We evaluated our system with 8 worker nodes and one specially configured \phubbox node. The workers were dual socket Broadwell Xeon E5-2680 v4 systems 
and 64 GB of memory using 8 dual-rank DDR-2400 DIMMs. Each worker had a GTX 1080 Ti GPU 
and one Mellanox ConnectX-3 InfiniBand card with 56 Gbps bandwidth in the same NUMA domain. The \phubbox machine was a dual socket Broadwell Xeon E5-2690 v4 system with 28 cores 
and 128 GBs of memory using 8 dual-rank DDR-2400 DIMMs. \phubbox had 10 Mellanox ConnectX-3 InfiniBand cards, with 5 connected to each socket. Hyperthreading was disabled. Machines were connected with a Mellanox SX6025 56 Gbps 36-port switch.

The machines ran CentOS 7.3 with CUDA 8 and CuDNN 7 installed. Our modifications to \mxnet and its PS (PS-Lite) were based on commit 2ce8b9a of the master branch in the PS-Lite repo. We built \mxnet with GCC 4.8 and configured it to use OpenBLAS and enable SSE, the Distributed Key Value Store, the \mxnet Profiler, and OpenMP. We used Jemalloc, as suggested by \mxnet.

\subsection{DNNs Used in the Evaluation}

We evaluated \phub{}'s performance by training state-of-the-art deep neural networks using reference code provided with \mxnet. 
We implemented cache-enabled SGD with Nesterov's accelerated gradient method \cite{nesterov1983method} and aggregator for \phub{}. We chose a per GPU batch size of 32 when possible; for ResNet 269 and ResNext 269, we used 16 and 8, respectively, since 32 did not fit in the GPU. We did not use MXNet's GPU memory optimizations~\cite{chen2016training} because they slow down training.

\begin{table}[t!]
	\centering
	\footnotesize
	\begin{tabular}{|c|c|c|c|}
		\hline 
		Name (Abbr)           & Model Size & Time/batch & Batch \\
		\hline
		AlexNet (AN)      & 194MB & 16ms &  32 \\
		\hline 
		VGG 11 (V11)       & 505MB & 121ms & 32 \\
		\hline
		VGG 19 (V19)      & 548MB & 268ms & 32 \\
		\hline
		GoogleNet (GN)   &  38MB & 100ms & 32 \\
		\hline
		Inception V3 (I3) & 91MB  & 225ms & 32 \\
		\hline
		ResNet 18 (RN18)   & 45MB & 54ms & 32 \\
		\hline
		ResNet 50 (RN50)   & 97MB & 161ms & 32 \\
		\hline  
		ResNet 269 (RN269)  & 390MB & 350ms & 16 \\
		\hline
		ResNext 269 (RX269) & 390MB & 386ms & 8 \\
		\hline
	\end{tabular}
	\caption{Neural networks used in our evaluation. Time/batch refers to the forward and backward compute times for a batch.}
	\label{table:networkCharacterization}
\end{table}

Table \ref{table:networkCharacterization} summarizes the neural networks used in our evaluation, which include both winners of the ImageNet challenge and other recent, popular networks. We used the reported model size from \mxnet and measured the forward and backward passes time on a single GPU. 

We report only training throughput in our evaluation since our modifications did not change accuracy because they did not change computations that were performed. We trained multiple DNNs to convergence to verify this.

\subsection{Training Performance Evaluation}
We include multiple results to highlight the effects of different software and hardware optimizations on \phub{}'s training performance. We measured training performance by comparing the total time of 200 iterations. We used two IB network configurations. This lets us compare training performance for two different compute/bandwidth ratios: (1) where GPUs were much faster than the network, and (2) with ample network bandwidth resources. In both setups, we used 8 workers.

\begin{figure}[t!]
	\includegraphics[width=\linewidth,trim=8 4 4 4,clip]{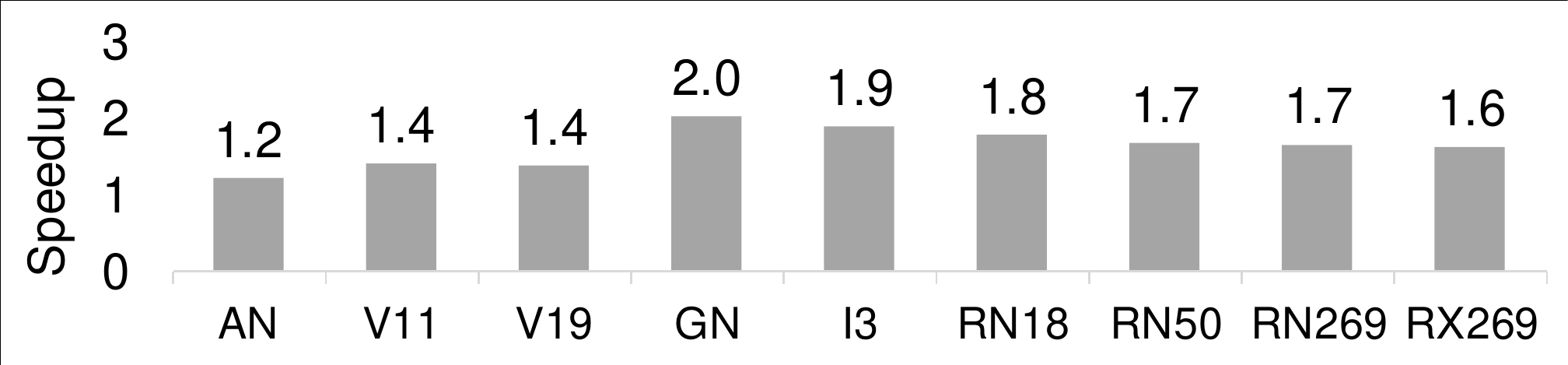}
	\caption{Speedup from a faster data plane that supports zero copy.}
	\label{fig:IBBenefits}
\end{figure}

\subsubsection{Benefit of a Faster Data Plane}
Figure \ref{fig:IBBenefits} shows the performance of replacing the communication stack of the \mxnet PS with a native InfiniBand implementation (\psliteib) that had all optimizations noted in \mysection\ref{sec:IBOptimization}. This lets us see the benefit of switching to an optimized network stack without changing the PS architecture. 
We used our \textit{enhanced baseline \psliteib} in all the following evaluation.

\subsubsection{Other Software and Hardware Optimizations}
We now quantify further benefits from \phub{}'s software and hardware optimizations. 
We used CS \psliteib in this comparison.
\pshard{} results were obtained by running \phub software on each worker as CS PSs. \phubbox results represent running \phub software on our single \phubbox machine as a NCC PS. We omit results for NCS and CC PSs for clarity. They performed similarly to \phubbox results.

\begin{figure*}[t!]
	\includegraphics[width=\linewidth,trim=4 4 4 4,clip]{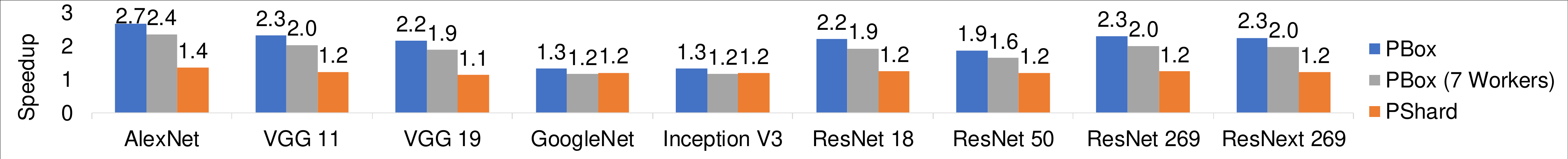}
	\caption{Training performance on a cloud-like 10 Gbps network. Results are normalized to sharded \psliteib (\textit{enhanced baseline}).}
	\label{fig:real-8gb}
\end{figure*}

\begin{figure}[t!]
	\includegraphics[width=\linewidth,trim=8 4 4 4,clip]{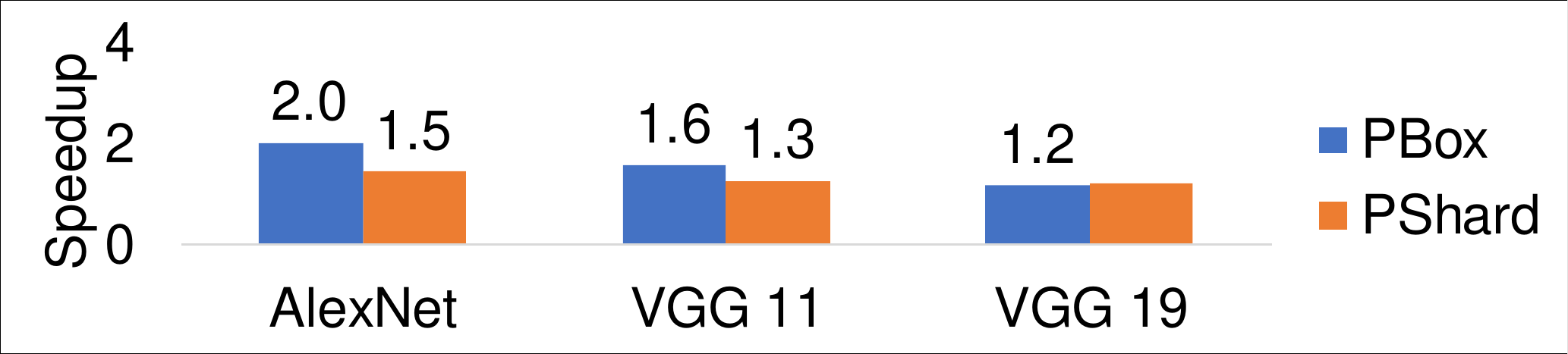}
	\caption{Training performance on a 56 Gbps network compared to \psliteib (\textit{enhanced baseline}). Computation speed bottlenecked training throughput for all but AlexNet and VGG.}
	\label{fig:real-56gb}
\end{figure}

\begin{figure}
	\includegraphics[width=\linewidth,trim=4 2 2 4,clip]{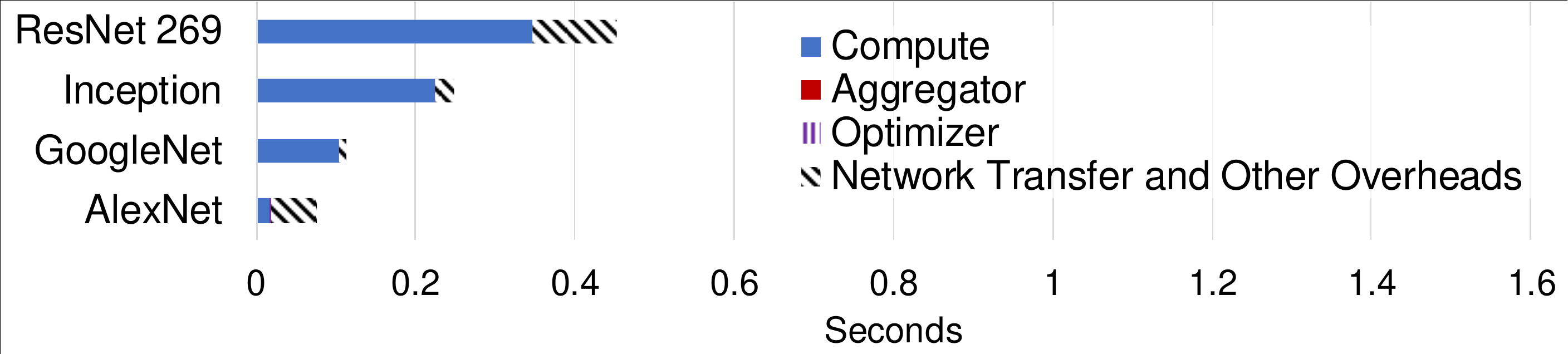}
	\caption{Progressive overhead breakdown of \phub. Compared to Figure~\ref{fig:overheadBreakdown}, GPU compute time now dominates training time. Aggregator and optimizer have minimum overhead, and are barely visible.}
	\label{fig:phubOverheadBreakdown}
\end{figure}

Figure~\ref{fig:real-8gb} shows training performance on a cloud-like 10 Gbps network, obtained by down-clocking our IB links. In this configuration, the ratio of GPU batch execution time to network serialization delays is such that the reduced communication and faster aggregation of \phubbox significantly affects runtime. In addition, we provide speedup when training with only 7 workers and \phubbox, \textit{so that the total machine count in the system is equal to the baseline}.

Figure~\ref{fig:real-56gb} shows training performance on 56 Gbps InfiniBand.
In this setup, for networks such as GoogleNet, Inception, ResNet, and ResNext, forward and backward pass execution time limits training throughput; thus, raising network bandwidth only marginally affects the total throughput of \phubbox versus \psliteib. Since \phub never slows down training, we omit results of these networks (1x speedup) for clarity. We expect larger speedups with newer, faster GPUs, such as the NVidia V100 for these networks. Significant speedup is still achieved with models that have large communication-to-computation ratios, such as AlexNet and VGG; these models remained network-bound even on 56 Gbps links.

The gap between \pshard and \psliteib signifies the benefit of software optimizations in \mysection\ref{sec:tallvswide}-\mysection\ref{sec:phub}, while the gap between \pshard and \phubbox highlights the benefit of both a non-colocated server that \textit{halve the per link bandwidth usage, which made a significant performance difference}, and the optimizations in \mysection\ref{sec:phub}.



Figure \ref{fig:phubOverheadBreakdown} breaks down the overhead in different distributed training stages when running \phub in the same setup as Figure \ref{fig:overheadBreakdown}. Compared to Figure \ref{fig:overheadBreakdown}, \textit{\phub reduces overheads from data copy, aggregation, optimization, and synchronization and fully overlaps these stages, shifting the training back to a compute-bound workload.}

\subsection{Performance with Infinitely Fast Compute}
We used a benchmark to assess the efficiency of \phub{}'s gradient processing pipeline to avoid being bottlenecked by our workers' GPUs. We implemented a special \mxnet engine, called \code{ZeroComputeEngine}, based on the original \code{ThreadedEnginePerDevice}, which replaces training operators (such as convolution) with an empty routine. Only the synchronization operators (\code{WaitForVar}, \code{KVStoreDistPush} and \code{KVStoreDistPull}) are actually executed. This engine effectively simulates arbitrarily fast forward and backward passes on the worker side, pushing the limits of \phub.

We used ResNet 18 as the test network. We first measured how fast each worker can run in this setup by running a single worker with the \phubbox, then gradually added more workers to plot total system throughput.

\begin{figure}[t!]
	\includegraphics[width=\linewidth,trim=1 1 1 1,clip]{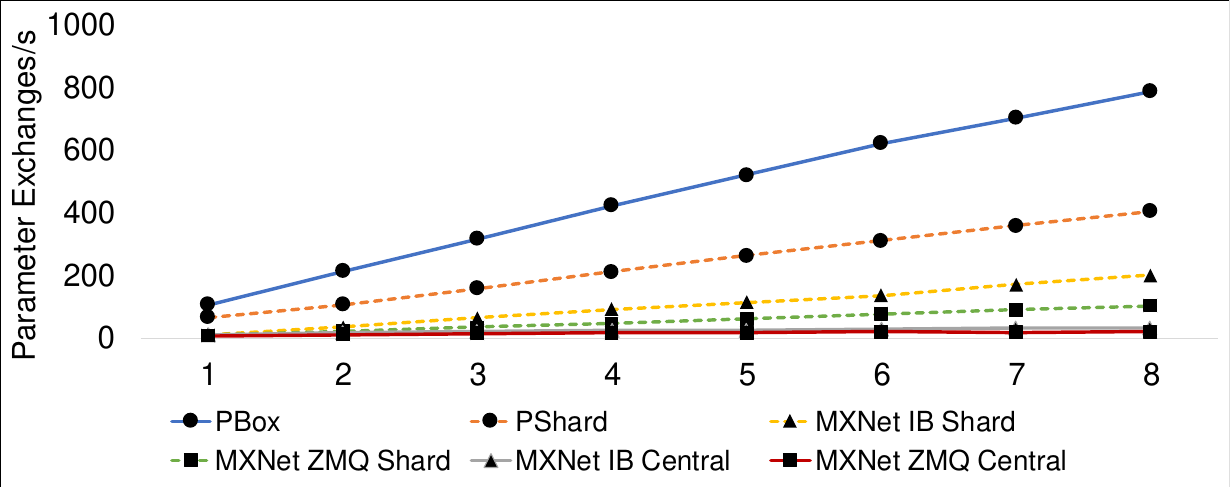}
	\caption{\phubbox provides linear scaling of throughput for 8 worker nodes with infinitely fast compute, training ResNet 18.}
	\label{fig:FakeTrainingWithAggOpt}
\end{figure}

Figure \ref{fig:FakeTrainingWithAggOpt} shows the results of running the benchmark with \phubbox, \pshard and multiple baseline configurations. \phubbox  provided linear scaling with 8 workers and outperformed the baseline by a large margin (up to 40x). \phubbox had 2x the speedup of \pshard{} because each of its interface needed to move only about half the amount of data compared to colocated servers.

\subsection{Exploiting Locality}
\label{sec:locality}
To postpone hitting the memory bandwidth limit, it is crucial to exploit locality in network interfaces and processor caches. This section evaluates the effectiveness of \phub{}'s key assignment scheme and tall aggregation/optimization in leveraging locality.

\vspace{0.05in}
\noindent \textbf{Key Affinity in \phubbox:}
\label{sec:affinity}
We evaluate two schemes for connecting workers to \phubbox to exploit locality and load balancing. In \textit{Key by Interface/Core mode}, workers partition their keys equally across different interfaces on the \phubbox. This mode better utilizes cache by binding a key to a particular interface, core and a NUMA node. 
This mode also exploits locality in time as workers are likely to generate the same key close to each other in synchronous training.

In \textit{Worker by Interface mode}, each worker communicates with the server through a single interface. This lets \phub exploit locality within a single worker. It also provides naturally perfect load balancing across interfaces and cores at the cost of additional communication and synchronization for each key within the server because keys are scattered across all interfaces and sockets.

We found that Key by Interface/Core provided 1.43x (790 vs 552 exchanges/s) better performance than Worker by Interface mode with \code{ZeroComputeEngine}. The locality within each worker could not compensate for synchronization and memory movement costs.

\vspace{0.05in}
\noindent \textbf{Tall vs. Wide Parallelism:}
We evaluated tall aggregation vs \mxnet{}'s wide approach with ResNet 50. Tall outperformed wide by 20x in terms of performance  and provides near-perfect scaling to multiple cores. Tall aggregation benefited from increased overlap compared to wide, and wide was further hurt by the cost of synchronization.


\vspace{0.05in}
\noindent \textbf{Caching Effectiveness in \phub:}
\label{eval:cache}
Caching benefits many \phub operations. For example, models can be sent directly from cache after being updated, and aggregation buffers can reside in cache near the cores doing aggregation for those keys. We now evaluate the effectiveness of caching in \phub by measuring memory bandwidth usage.

\begin{table}[t!]
	\centering
	\footnotesize
	\begin{tabular}{|c|c|c|}
		\hline 
		& Mem BW & Throughput\\
		\hline
		Opt/Agg Off & 77.5 & 72.08 \\
		\hline 
		Caching Opt/Agg & 83.5 & 71.6 \\
		\hline
		Cache-bypassed Opt/Agg & 119.7 & 40.48 \\
		\hline
	\end{tabular}
	\caption{Bidirectional memory bandwidth (GB/s) utilization in \phub when training VGG with 8 workers. The maximum memory bandwidth for the machine is 137 GB/s for read-only workloads and 120 GB/s for 1:1 read:write workloads as measured by LikWid and Intel MLC.}
	\label{table:cacheUtilizationAndAggregationOverhead}
\end{table}

Table \ref{table:cacheUtilizationAndAggregationOverhead} shows the memory bandwidth costs of communication, aggregation, and optimization on \phubbox. We used 8 workers running a communication-only benchmark based on the VGG network, chosen because it had the largest model size. We first ran the benchmark with no aggregation or optimization, and we then added our two aggregation and optimization implementations.

Without aggregation and optimization, \phubbox{}'s bidirectional memory bandwidth usage was stable at 77.5 GB/s. No cache was used in this case because \phubbox did not touch the data (only the network interface did).

We found that the caching version of the aggregator and optimizer performed significantly better than the cache-bypassing version, which hit the maximum memory bandwidth available on the \phub machine when combined with the memory bandwidth of worker sends and receives. The caching version, on the other hand, added only 8\% to total memory bandwidth usage; aggregation and optimization added only 1\% of overhead to the overall throughput in this benchmark, fully overlapping gradient transfer.

\subsection{Tradeoffs in Fine-Grained Key Chunking}
\label{sec:commParam}
We now examine tradeoffs in the communication layer concerning the size of key chunks and queue pair counts.

\begin{figure}[t!]
	\includegraphics[width=\columnwidth,trim=4 8 2 4,clip,]{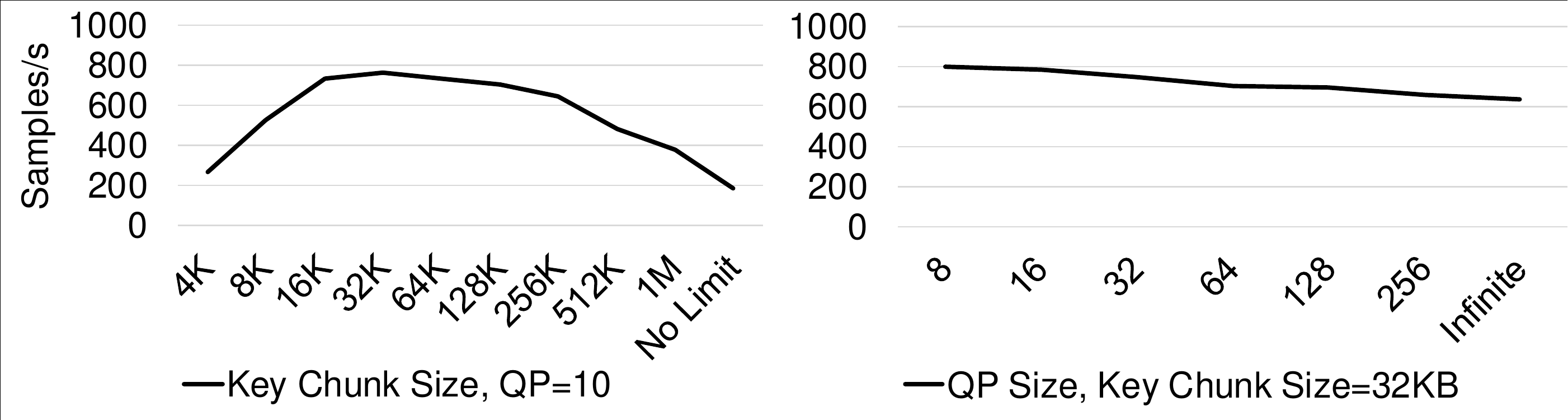}
	\caption{Effect of chunk size and queue pair count on throughput.}
	\label{fig:QPandKeyChunk}
\end{figure}

\vspace{0.05in}
\noindent \textbf{Size of key chunks:}
\label{sec:keyChunking}
\phub leverages fine-grained key chunking to better balance load and overlap gradient reception and aggregation. Figure \ref{fig:QPandKeyChunk} (left) evaluates the effect of chunk size with \code{ZeroComputeEngine} on \phubbox. Larger chunk sizes improved network utilization, while smaller sizes improved overlapping. We found 32KB chunk size to be optimum: this is likely due to our interfaces' maximum injection rate and aggregation pipeline latency.

\vspace{0.05in}
\noindent \textbf{Queue Pair Count:}
A worker needs at least one queue pair per interface with which it communicates. Queue pairs have state, which is cached on the card. When that cache misses frequently, communication slows. For \phubbox to use 10 interfaces, we need a minimum of 10 queue pairs per worker. More queue pairs could enable concurrent transmission from the same worker and reduce head of line blocking, but it increases the queue pair cache miss rate. Figure \ref{fig:QPandKeyChunk} (right) evaluates the tradeoff, showing that fewer queue pairs was optimal.

\subsection{Limits on Scalability}
\label{sec:scability} 
The scalability of \phub is inherently limited by available total memory, network or PCIe bandwidth. This section explores how close \phub gets to these limits. We use \phubbox to answer these questions. \phubbox achieves a 1:1 read:write memory bandwidth of 120 GB/s and a bidirectional network bandwidth of 140 GB/s. To determine how much bandwidth can be utilized,
we added an additional IB interface to each of our 8 machines to emulate 16 workers and configured varying numbers of emulated workers running \code{ib\_write\_bw},
each with 10 QP connections to the \code{ib\_write\_bw} process on \phubbox. These pairs of processes did repeated RDMA-writes to two 1 MB buffers on the other side. We set the PCIe read request size to 512 bytes. This configuration was chosen to mirror the setup of an actual training system while maximizing total system throughput. 

To our surprise, we found that the peak memory bandwidth usage never exceeded more than 90 GB/s, far from the limit of both the aggregate network card and memory. This suggests that the bottleneck lies somewhere else. 

We then built a loopback microbenchmark that used the IB cards to copy data locally between RDMA buffers. This isolated the system from network bottlenecks and involved only the NIC's DMA controllers and the processor's PCIe-to-memory-system bridge. This microbenchmark also achieved only 90 GB/s. Based on this experiment, we believe that \textit{the limit of throughput in our current \phub system is the PCIe-to-memory-system bridge.}

\begin{figure}[t!]
	\includegraphics[width=\linewidth,trim=2 2 2 2,clip]{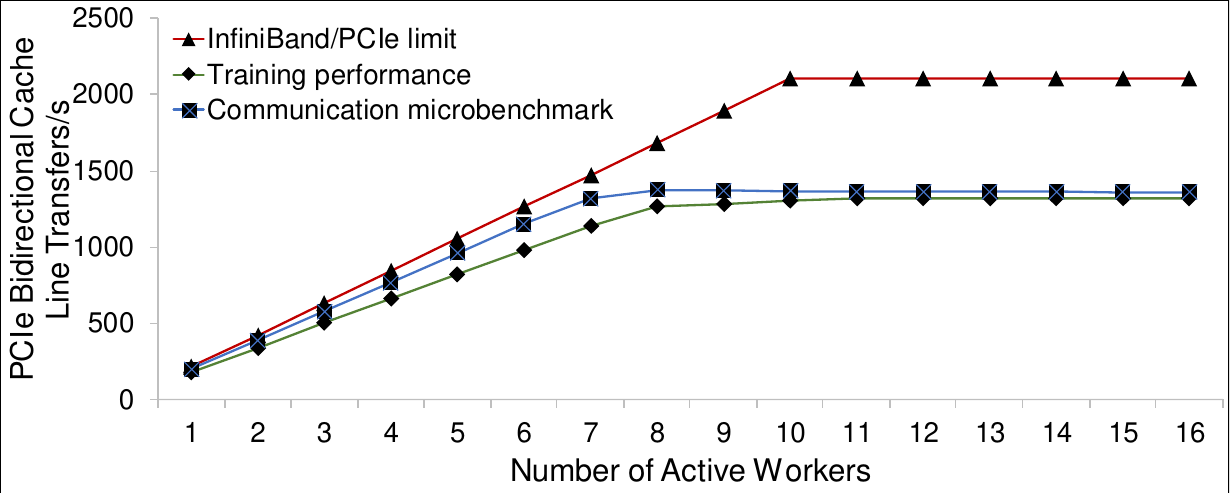}
	\caption{\phubbox{} scalability is limited by the throughput of the PCIe to the on-chip network bridge of the PBox processors. \phub{} can utilize 97\% of the measured peak PCIe bandwidth.}
	\label{fig:scalablity}
\end{figure}

Figure \ref{fig:scalablity} summarizes this experiment. The InfiniBand/PCIe limit line shows an ideal case where unlimited cache line transfers can be performed. However, this rate was not achievable even with a microbenchmark, which poses a hard upper bound on how fast \phub can run during training. We also see that, when training VGG with \code{ZeroComputeEngine}, as more workers are added, \phubbox{}'s performance approached the microbenchmarks (97\%), demonstrating \phub{}'s ability to fully utilize system resources. The gap in the plot between \phubbox and the microbenchmark 
is due to the overhead of scheduling operations in \mxnet and straggler effects in workers. \phubbox{} hit the limit at a sustained 80GB/s memory bandwidth.

In real training, however, \phubbox{}'s scalability limit was difficult to reach. Recent work (\cite{keskar2016large, lecun1524efficient}) describes the difficulty of generalization with large batch sizes; it is not advantageous to blindly scale deep learning to a large number of workers without considering statistical efficiency~\cite{youspeeding, koliousiscrossbow}. One example \cite{ImageNetIn1Hour} reports that ResNet 50's statistical efficiency drops with aggregate batch sizes larger than 8192 on a system with 256 GPUs on 32 machines (with a mini-batch size of 32 per GPU). To assess whether \phubbox could reach this scale, we measured the memory bandwidth usage for ResNet 50 with 8 workers using the same batch size. We found that \phubbox required only 6GB/s memory bandwidth and an aggregated 4GB/s network bandwidth. This suggests that our \phubbox prototype could scale to rack-level and support up to 120 worker machines training this network. In other words, our prototype could support sufficient scalability to handle cutting-edge training tasks.

On the other hand, the scalability bottleneck (PCIe controller) in our current prototype is specific to this particular platform, but it can change. For example, recently released AMD Epyc~\cite{AMDEpyc} processors provide nearly triple the Stream Triad performance
(290 GB/s)~\cite{EpycBenchmark} and 40\% more PCIe bandwidth than our
\phubbox machine. We would expect Epyc to support 40\% more
throughput.

\subsection{Effectiveness of \phubbox as a Rack-Scale Service}
\label{sec:hierarchicalEval}
We now evaluate effectiveness of \phubbox as a rack-scale service with two typical scenarios in a 10 Gbps cloud-like environment: (1) when multiple jobs are training in parallel in a rack, sharing the same \phubbox instance with different key namespaces and (2) when a training job crosses rack boundaries, and \phub performs hierarchical reduction.

\begin{figure}[t!]
	\includegraphics[width=\linewidth,trim=2 2 2 2,clip]{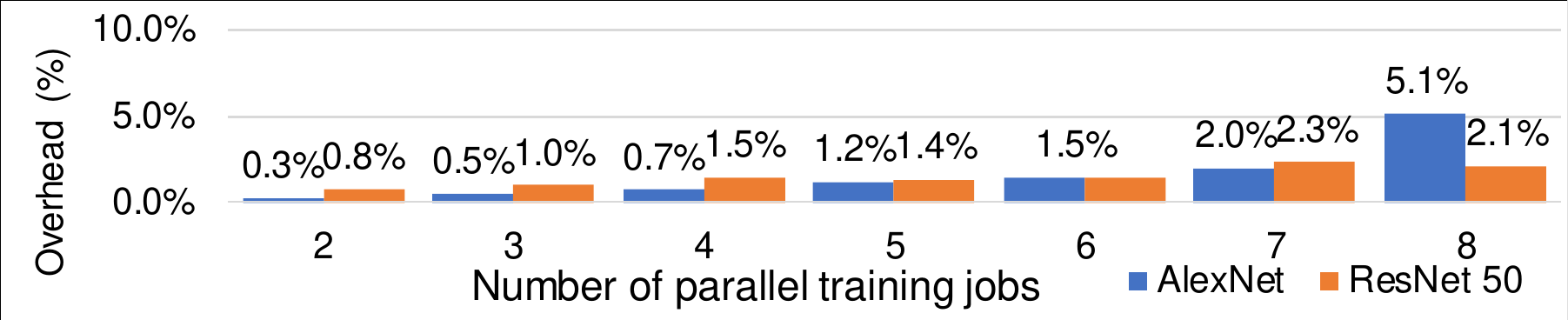}
	\caption{Overhead of multiple parallel training jobs sharing the same \phubbox instance.}
	\label{fig:multijobs}
\end{figure}

Figure \ref{fig:multijobs} shows the overhead of running multiple independent training jobs when sharing a single \phubbox instance. AlexNet saw a 5\% drop in per-job throughput when running 8 jobs, likely due to frequent invocation of optimizer and less effective caching; ResNet 50 saw a smaller impact as it is compute bound.

\begin{figure}[t!]
	\includegraphics[width=\linewidth,trim=2 6 2 2,clip]{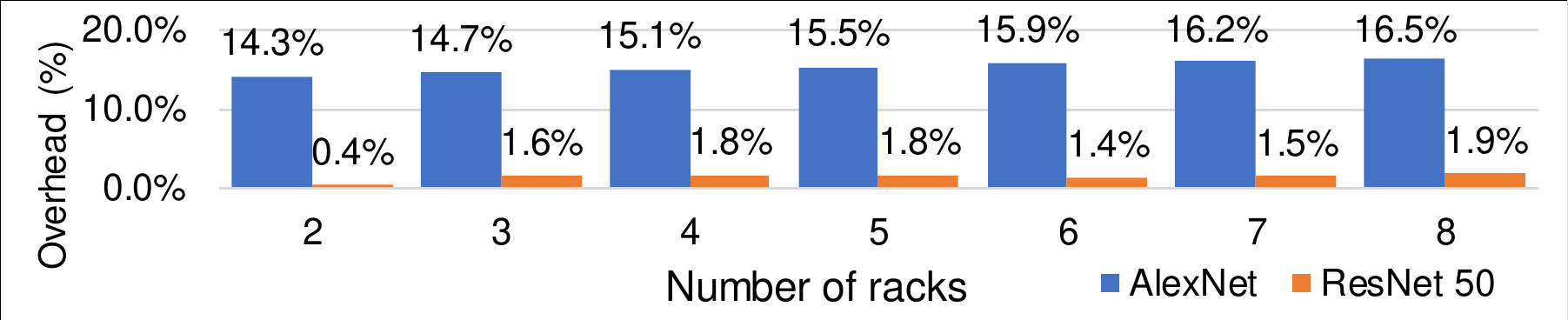}
	\caption{Emulated overhead of hierarchical reduction with \phubbox.}
	\label{fig:hierarchical}
\end{figure}

Figure \ref{fig:hierarchical} emulates the a single cloud-based training job whose VMs span N racks, and each rack contains 8 workers and 1 \phubbox. The \phubbox uses a widely used ring reduction algorithm~\cite{baidures3:online,DBLP:journals/corr/abs-1802-05799} for inter-rack aggregation. 

Since we have only one \phubbox machine, we model this ring reduction by sending and receiving N chunk-size messages sequentially, each performing one additional aggregation, for each of the keys, after local rack has finished aggregation. We assume each rack would finish its local aggregation at roughly the same time, as stragglers can exist regardless of rack assignment. Therefore, this faithfully estimates overhead of \phub{}'s hierarchical reduction.

The lost throughput of AlexNet comes from added latency of multiple rounds of communication, but is compensated by drastically reduced cross-rack communication, and thus we would expect speedup in real deployment. On the other hand, we again observed virtually no loss of throughput in ResNet 50.

\subsection{Rack-scale cost model}

Is a cluster built with PHubs and a slow network more cost effective than one with sharded PSs and a fast network? This section explores this question using a simple cost model. We consider the cost of three cluster components: worker nodes, PHub nodes, and network gear. We use advertised prices from the Internet; while a datacenter operator might pay less, the ratios between component prices should still be similar. The baseline is a cluster running MXNet IB with colocated sharded PSs; we compare this to a PHub deployment in terms of throughput per dollar.

The model works by computing the cost of a worker node, and adding to it the amortized cost of its network usage; for the PHub deployment, it also includes the amortized cost of the worker's PHub usage. This allows us to compare the cost of worker nodes in deployments with different numbers of workers per rack, switch, or PHub. We capture only the most significant cost, and include only capital cost, since operational costs are dominated by GPU power usage and thus differences would be small.

We model a standard three-layer datacenter network with some simplifying assumptions: racks hold as many machines as may be connected to a single switch, all switches and cables are identical, and oversubscription happens only at ToR switches. We model network costs by charging each worker the NIC per-port cost $N$, the amortized cost of one ToR switch port $S$ and cable $C$, and fractional costs of ToR/aggregation/core switch ports and cables depending on the oversubscription factor $F$. Thus, the amortized cost of the network per machine is $A=(N+S+C)+F(4S+2C)$. 

Since our goal is to model costs for future deployments, we make two changes from our experimental setup. Instead of 10Gb IB, we use 25 Gb Ethernet. Instead of NVIDIA 1080 Ti's, we assume a future, faster GPU with similar cost $G$, but performance like today's V100. Based on the data in Figure~\ref{fig:gpuPower}, this keeps the compute/communication ratio similar to that of our experiments. We use ResNet-50 for comparison; we use our 10Gb IB results for the PHub setting and downclocked 40Gb IB for the MXNet IB baseline. We include 2\% overhead in the PHub numbers to account for aggregation between racks.



Workers are the same as in our evaluation, but with 4 GPUs. The cost $W$ is \$4117~\cite{worker-price} without GPUs; the GPU price $G$ is (\$699~\cite{nvidia-1080ti}). The 100Gb baseline uses Mellanox ConnectX-4 EN cards (\$795~\cite{mellanox-eth}) and 2m cables (\$94~\cite{mellanox-cable}). The 25Gb PHub workers use Mellanox ConnectX-4 Lx EN cards (\$260~\cite{mellanox-eth}) and 4-to-1 breakout cables (\$31.25 per port~\cite{mellanox-cable}).
The PHub node (also same as evaluation) cost $H$ is \$8407~\cite{phub-price}, plus 10 dual-port 25Gb Mellanox ConnectX-4 Lx EN cards (\$162.5 per port~\cite{mellanox-eth}). The cost of each baseline worker is $W+N+4G+A$, and the cost of a PHub worker is $W+N+4G+A+KP$, where $KP$ is the amortized PHub cost ($P=W+20N+20A$; $K$ is the worker to \phub ratio).

We use the Arista 7060CX-32S 32-port 100Gb Ethernet switch (\$21077~\cite{arista-price}) in both configurations, with breakout cables to connect 25Gb hosts.
With no oversubscription, each switch supports 16 100Gb baseline workers, or a PHub and 44 25Gb workers. With 2:1 oversubscription each switch could support a \phub and 65 25Gb workers; with 3:1, 76.


\begin{table}[tb!]
  \centering
  \footnotesize
  \begin{tabular}{|r|c|c|c|}
    \hline
    & \multicolumn{3}{c|}{Throughput/\$1000} \\
                                 & Future GPUs & Spendy & Cheap \\
    \hline
    100Gb Sharded 1:1 &             46.11 &          14.57 &      60.41\\
    \hline 
    25Gb PHub     1:1 &             55.19 &          15.30 &      77.21\\
    \hline 
    25Gb PHub     2:1 &             57.71 &          15.49 &      82.24\\
    \hline 
    25Gb PHub     3:1 &             59.03 &          15.58 &      84.95\\
    \hline
  \end{tabular}
  \caption{Datacenter cost model comparing 25GbE PHub deployments with 100GbE MXNet IB on ResNet-50. Higher is better. The Future GPU PHub deployment with 2:1 oversubscription provides 25\% better throughput per dollar.}
  \label{table:costModel}
\end{table}


Table~\ref{table:costModel} compares a full-bisection-bandwidth 100GbE sharded MXNet IB deployment with 25GbE PHub deployments with varying oversubscription. With 2:1 oversubscription, the PHub deployment provides 26\% better throughput per dollar. We consider two other configurations: first, a ``lower bound'' using today's expensive V100's, where the 2:1 PHub deployment provides only 6\% improvement; and a ``GPU-focused'' one using cheap CPUs (E5-2603 v4) in workers, providing 36\% improvement.

\section{Additional Related Work}
\label{sec:discussion}

This section augments the related work discussed in \mysection\ref{sec:background}.

\begin{figure}[t!]
	\includegraphics[width=\linewidth,trim=8 4 4 2,clip]{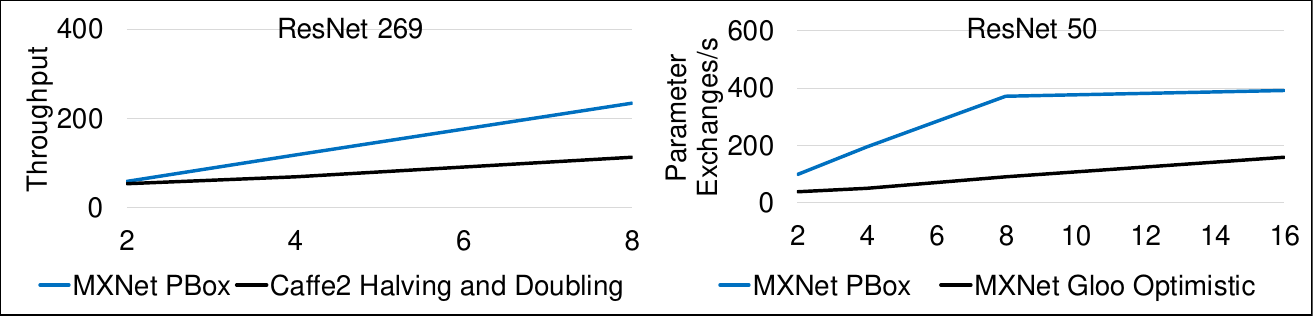}
	\caption{Left: Comparing Caffe2 + Gloo and \mxnet + \phubbox on an 10Gbps InfiniBand network. Right: Comparing \mxnet + Gloo and \mxnet + \phubbox on a 56Gbps InfiniBand network with \code{ZeroComputeEngine}. }
	\label{fig:gloo}
\end{figure}

\vspace{0.05in}
\noindent \textbf{Other Communication Schemes:}
Parameter servers are not the only way to perform model updates. Frameworks such as CNTK and Caffe2 can use HPC-like approaches, such as collective communication operations~\cite{Thakur:2005:OCC:2747766.2747771, firecaffe}.

To understand how \phub compares to other communication schemes, we first ran Caffe2 and \mxnet with \phubbox on a cloud-like network. We used InfiniBand for both systems. We evaluated the fastest algorithm in Gloo: recursive halving and doubling, used in \cite{ImageNetIn1Hour}. Figure \ref{fig:gloo} (left) shows the result: \phubbox was nearly 2x faster. 

We ported Gloo to \mxnet to better asses both systems. Gloo implements blocking collective operations, but \mxnet expects non-blocking operations. Therefore, we measured an optimistic upper bound by letting Gloo start aggregating the entire model as soon as the backward pass started, as if all gradients were available instantaneously. Since Gloo only does reduction, we ran our SGD/Nesterov optimizer on all nodes after reduction was complete. We used 56 Gbps IB and \code{ZeroComputeEngine} to remove network and worker bottlenecks. Figure \ref{fig:gloo} (right) shows the result; 
\phubbox sustained higher throughput and provided better scaling up to its limit. Two reasons account for this difference. First, collectives suffer from the same problem as colocated PSs: the interface on each participating node must process nearly 2x the data (Gloo's \code{allreduce} starts with a \code{reduce-scatter} followed by an \code{allgather}~\cite{Thakur:2005:OCC:2747766.2747771}). 
Second, collectives frequently use multi-round communication schemes 
($logN$ rounds for $N$ workers in this case), whereas \phubbox uses only 1 round. 


\vspace{0.05in}
\noindent \textbf{Compression, Quantization, Sparse Vector Communication, and Other Mechanism for Traffic Reduction:}
Orthogonal to our work are techniques to reduce gradient traffic. These techniques trade higher overhead in preparing and processing network data for lower network bandwidth usage. For example, \mxnet{} supports a 2-bit compression scheme, similar to \cite{cntk1bt}. We compared \phub running on \phubbox to \psliteib with 2-bit compression: \phubbox without compression still beat \psliteib by 2x. 


Other examples include Sufficient Factor Broadcast (SFB) \cite{poseidon,DBLP:journals/corr/XieKX14}, which decomposes the gradient of a fully connected layer (FCL) into the outer product of two vectors. SFB uses a P2P broadcast scheme whose overhead scales quadratically with the number of machines, making it suboptimal for large scale training. Project Adam~\cite{projectAdam} sends activation and error gradient vectors for reconstruction on server. Both techniques have limited applicability as they only apply to FCLs, which are small or unused in recent neural networks~\cite{DBLP:journals/corr/SzegedyIV16,RESNET,inception-v3,ResNext}.

\phub can also work with gradient compression \cite{DBLP:journals/corr/abs-1712-01887} to gain further benefits from its low latency communication stack, fast aggregation and optimization.



\section{Conclusion}
\label{sec:conclusion}
We found that inefficient PS software architecture and network environment-induced overhead were the major bottlenecks of distributed training with modern GPUs in the cloud, making DDNN training a communication-bound workload. To eliminate these bottlenecks, we proposed \phub, a high performance multi-tenant, rack-scale PS design, with co-designed software and hardware to accelerate rack-level and hierarchical cross-rack parameter exchange. Our evaluation showed that \phub provides up to 2.7x higher throughput, with 25\% better throughput per dollar.




\bibliographystyle{plain}
\bibliography{paper}
	

	
\end{document}